\documentclass[aip,apl,reprint]{revtex4-1}
\usepackage{dcolumn}%
\usepackage{amssymb}
\usepackage{graphicx,amsmath}
\usepackage{bm}
\usepackage{times}
\usepackage{float}
\begin{document}

\title{ Parametric amplification by coupled flux qubits.}

\author{M. Reh\'{a}k}
\author{P. Neilinger}
\author{M. Grajcar}
\affiliation{Department of Experimental Physics, Comenius University, SK-84248 Bratislava, Slovakia}
\affiliation{Institute of Physics, Slovak Academy of Science, Bratislava, Slovakia}
\author{G. Oelsner}
\author{U. H\"{u}bner}
\author{E. Il'ichev}
\altaffiliation{Novosibirsk State Technical University, 20 K. Marx Ave., 630092 Novosibirsk, Russia}
\author{H.-G. Meyer}
\affiliation{Institute of Photonic Technology, P.O. Box 100239, D-07702 Jena, Germany}
\begin{abstract}
 We report the parametric amplification of a microwave signal in a Kerr medium formed from superconducting qubits. Two mutually coupled flux qubits, embedded in the current antinode of a superconducting coplanar waveguide resonator, are used as a nonlinear element. Shared Josephson junctions provide the qubit-resonator coupling, resulting in a device with a measured gain of about 20 dB. We argue, that this arrangement represents a unit cell which can be straightforwardly extended to a quasi one-dimensional quantum metamaterial with a large tunable Kerr nonlinearity.
\end{abstract}

\maketitle


New developments in circuit quantum electrodynamics has resulted in the possibility of signal detection close to and even below the standard quantum limit. This work was motivated by microwave quantum engineering,\cite{You03, Grajcar05a,Xiang13} including quantum information processing devices.\cite{Reed12,Fedorov12}
The sensitivity of cryogenic semiconductor
amplifiers with reasonable power consumption in both the MHz \cite{Oukhanski03} and GHz (commercially available) range are all currently above the quantum limit. At very low temperatures it is quite natural to use the parametric effect for amplification which adds no additional noise to the signal. In practice, a nonlinear superconducting oscillator can be used for this purpose.\cite{Kuzmin83, Zagoskin08}
In this case, usually a nonlinearity of
superconducting weak links \cite{Beltran08, Yamamoto08} is exploited.  Very recently the
squeezing of quantum noise and its measurement below the standard quantum limit was
demonstrated.\cite{Mallet11}  These successful experiments have motivated the
development of new parametric amplifiers based on
Josephson junctions,\cite{Bergeal10, Hatridge11}  DC
squids,\cite{Abdo09, Hatridge11,Sudqvist13} high kinetic inductance of weak links,\cite{Tholen07} or disordered superconductors.\cite{Eom12}
Superconducting qubits also exhibit a strong nonlinearity \cite{Ilichev07} and therefore are good candidates for parameteric amplification.\cite{Savelev12} 

\begin{figure}
\centerline{\includegraphics[width=7cm]{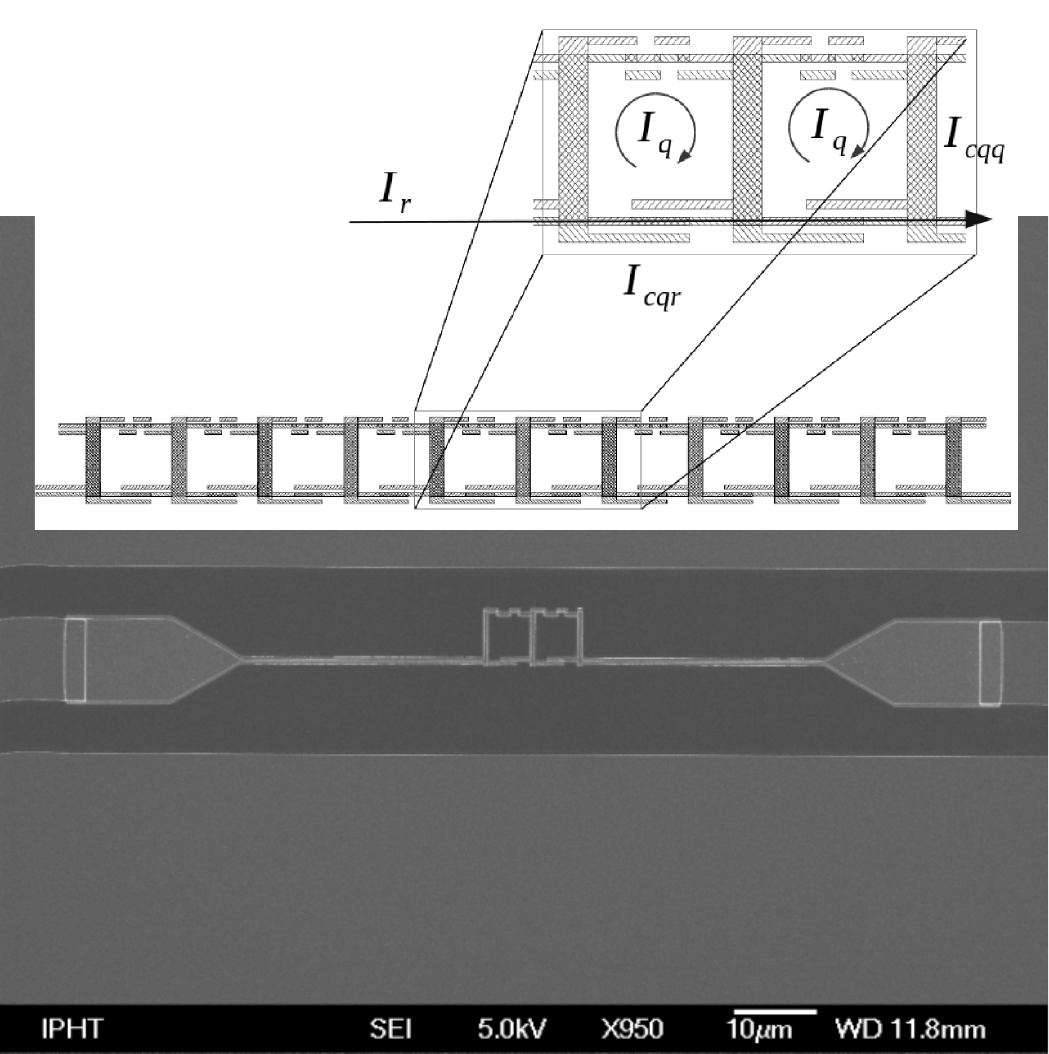}}
	\caption{Design of a unit cell which consists of two coupled qubits and its proposed extension to an array of qubits (upper part). The bottom and upper electrodes are fabricated by the shadow evaporation technique and marked by different hatching. The Josephson
junctions are formed in areas where the hatchings are overlap. The qubits share a large Josephson junction providing ferromagnetic coupling. Each qubit is strongly coupled to the resonator by an Josephson junction. The SEM image of the sample (lower part) shows the middle part of the resonator with a unit cell of two qubits.}
\label{fig:cell}
\end{figure}

In this paper, we demonstrate parametric amplification exploiting the nonlinearity of a pair of superconducting flux qubits coupled to a coplanar waveguide resonator. These qubits represent a unit cell which can be easily extended to a
one-dimensional array. Such an array can be considered as a medium with large
Kerr nonlinearity.
\begin{figure*}
\includegraphics[width=6.5cm]{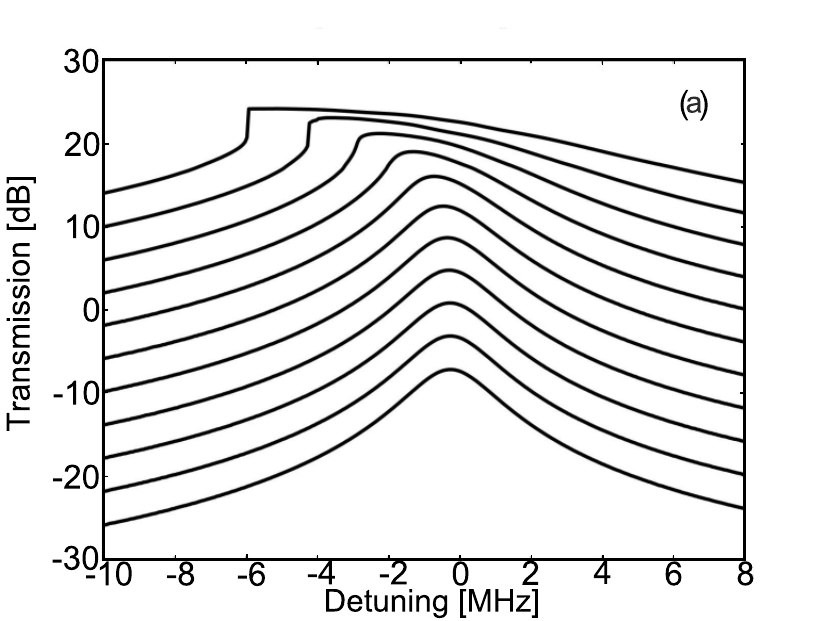}
\includegraphics[width=6.5cm]{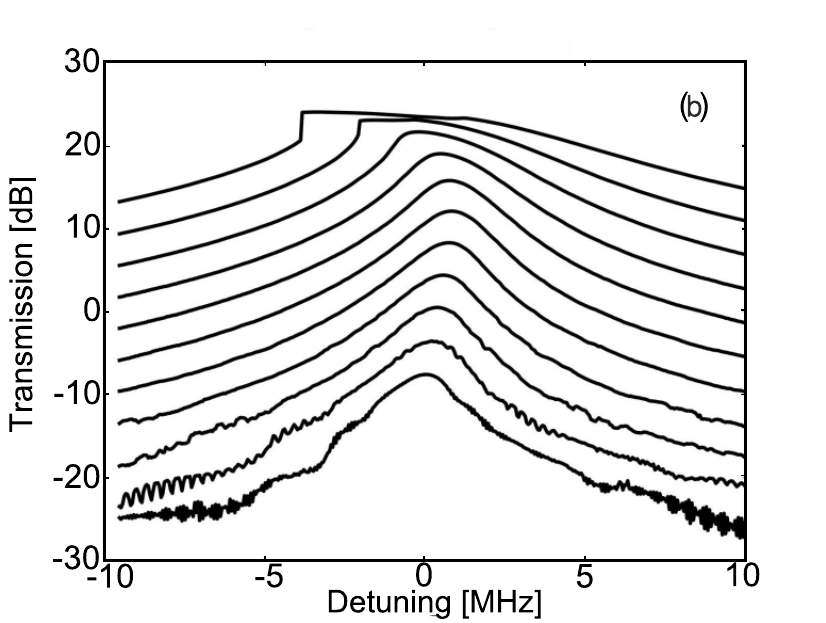}
\caption{ Transmission of the resonator measured around the third harmonic (7.45 GHz) for an applied magnetic flux $\Phi=0$ (a) and
$\Phi=\Phi_0/2$ (b). Curves, from bottom to top, correspond to an input power of -90 to -46 dBm in 4 dBm steps. The curves are shifted by an increment of 4 dB along the $y$-axes for better visibility.
\label{fig:nonlinear}}
\end{figure*}
Neglecting the interactions between the modes, a nonlinear oscillator can be described by the
Hamiltonian\cite{Yurke06}
\begin{equation}
H_n = \hbar \omega_n a^{\dagger}a+\frac{\hbar}{2}K_na^{\dagger}a^{\dagger}aa
\label{eq:Hn}
\end{equation}
where $\hbar$ is the reduced Planck constant, $a$ and $a^{\dagger}$ are the creation and the annihilation operators, $\omega_n$ is the angular frequency of the $n$-th mode of the oscillator, and
$K_n$ is the Kerr constant, which is a measure of the nonlinearity in the system.
The qubits contribute to the nonlinear inductance of the resonator per unit length according to
\begin{equation}
\tilde{L_r}(I_r)=\tilde{L}_0+\tilde{L}_2(I_r/I_{cqr})^2+\tilde{L}_4(I_r/I_{cqr})^4+\dots \;.
\label{eq:L}
\end{equation}
Then, the Kerr constant can be calculated by\cite{Yurke06}
\begin{equation}
K_n= -\frac{\hbar \omega_n^2}{{I^2_{cqr}}} \int_0^l dx u_n^4(x) \tilde{L}_2 ,
\label{eq:Kn}
\end{equation}
where $u_n(x) = \sqrt{\frac{2}{\tilde{L}_r l}}$sin$(\frac{n\pi x}{l})$, $l$ is the length of the resonator, $\tilde{L}_r$ the linear inductance of the resonator per unit length, $I_r$ the current flowing in the resonator, $I_{cqr}$ the critical current of the Josephson junction shared by
the resonator and qubit, and $\tilde{L}_i$ are the Taylor coefficients characterizing the nonlinearity of the  inductance.
The Kerr constant can be determined for every nonlinear medium.
In optics, the Kerr constants are very small while quantum information processing schemes require
large ones. Even for a single superconducting qubit the Kerr constant is not small. It may be further enhanced by using ferromagnetically coupled qubits \cite{Grajcar06, Smirnov03}. Moreover this constant
can be tuned over a wide range, even from positive to negative large values.

To demonstrate these features we fabricated two ferromagnetically coupled qubits embedded in a coplanar waveguide resonator. The niobium resonator is fabricated by e-beam
lithography and dry etching the 200 nm thick film on a silicon substrate. The aluminium qubit
structures are placed in the middle of the resonator and were prepared by the shadow evaporation
technique (figure~\ref{fig:cell}). To assure strong coupling between the resonator and the qubits (there are 6 Josephson junctions in each qubit), one of the Josephson junctions of each qubit, with critical current $I_{cqr}$, is integrated within the centerline of the resonator.
A bandwidth of the parametric amplifier is rather narrow, limited by a bandwidth of the resonator. 
However, the unit cell of two qubits can be extended to an array, forming a metamaterial in a coplanar waveguide (see figure~\ref{fig:cell}), where 
the enhanced interaction of the microwaves with one unit cell  caused by the multiple reflections of the microwave signal in the coupling capacitors of the resonator
is substituted by the interaction with a large number of
cells. By making use of such microwave waveguide instead
of a resonator the bandwidth can be increased 
considerably.\cite{Eom12}
The qubits in the array are weakly coupled by a shared large Josephson junction with a critical current $I_{cqq}$.
The coupling energy between adjacent qubits is \cite{Grajcar05}
\begin{equation}
J_{qq}=\frac{\Phi_0I_{qi-1}I_{qi}}{I_{cqq}}
\label{eq:Jqq}
\end{equation}
where $I_{qi}$ is the persistent current of the $i$-th qubit, $I_{cqq}$ is the critical current of the coupling Josephson junction shared by adjacent qubits, and  $\Phi_0$ is the magnetic flux quantum.
In the ground state the persistent currents in adjacent qubits flow through the coupling Josephson junction in the same direction in order to minimize the total energy of the two coupled qubits. By clever design
one can use the shadow evaporation technique to twist the electrodes of the coupling Josephson junction in such a way that ferromagnetic coupling is achieved (see figure~\ref{fig:cell}).
For one qubit centered in a $\lambda/2$ resonator the nonlinear inductance per unit length can be calculated from the relation \cite{Green_b02}
\begin{equation}
\tilde{L} = \frac{F}{4\pi^2}\frac{\Phi_0^2}{\Delta}\frac{I_q^2}{I_{cqr}^2}\delta(x-\frac{l}{2})
\label{eq:L_}
\end{equation}
where $\delta(x)$ is the delta function, $\Delta$ and $I_q$ are the tunneling amplitude and the persistent current of the superconducting flux qubit, respectively, and the function $F$ is defined as
\begin{equation}
F = \frac{1}{\pi}\int_0^{2\pi} d\phi \frac{cos^2\phi}{[1+\eta^2(f_x+\gamma\sin\phi)^2]^{3/2}},
\label{eq:F}
\end{equation}
where
$\gamma = \Phi_0I_qI_r/(2\pi I_{cqr}\Delta)$ and $\eta=\Phi_0I_q/\Delta$.
The coefficient $\tilde{L}_2$ is obtained from the Taylor expansion of the nonlinear inductance $\tilde{L}$.

Interactions with the measurement apparatus and the dissipative environment are modelled by making use of a standard model in the microwave region - by making use of the test and the dissipative ports (details see in Ref.~\onlinecite{Yurke06} and references therein).
Parametric amplification is achieved by strong pumping of the resonator at a
frequency $\omega_p/2\pi$.
The pump frequency is mixed with a weak signal of frequency $\omega_s/2\pi$ by the
nonlinear element, producing additional signals with angular frequencies $\omega_s$ and $2\omega_p-\omega_s$ (idler signal). In other words, two photons from the pump with angular frequency $\omega_p$ transform into two photons with angular frequencies
$\omega_s$ and $2\omega_p-\omega_s$, while the energy is conserved.
For strong pumping and in absence of fluctuations the system can be described by the classical
variables $\alpha=\langle a\rangle$, $\alpha^*=\langle a^{\dagger}\rangle$, and the
average number of photons in the resonator\cite{Yurke06}
$\bar{n}=\alpha\alpha^*$
\begin{eqnarray}
\bar{n}^3&+&\frac{2(\omega_0-\omega_p)K_n+2\gamma\gamma_3}{K_n^2+\gamma^2_3}\bar{n}^2+
\frac{(\omega_0-\omega_p)^2+\gamma^2}{K_n^2+\gamma^2_3}\bar{n} \nonumber \\
&=&\frac{2\gamma^2_1}{K_n^2+\gamma^2_3}\bar{n}_{in}
\label{eq:n}
\end{eqnarray}
where $\bar{n}_{in}$ is the average number of photons coming to resonator from input port in a
time interval $1/\gamma_1$ .
The amplitude of the transmission coefficient of the resonator, which can be measured directly by a network analyzer and is given by the ratio $|t|=\bar{n}/\bar{n}_{in}$
\begin{figure}
\includegraphics[width=6.5cm]{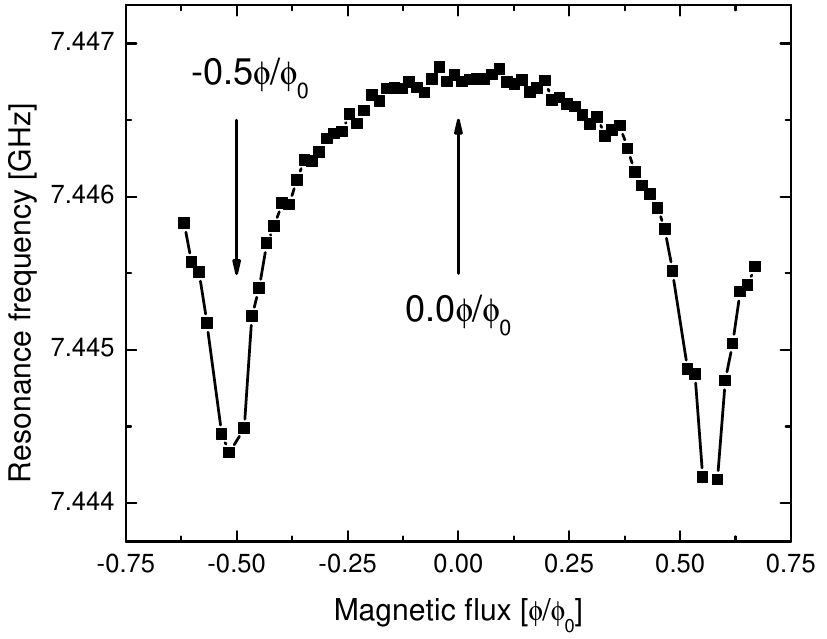}
	\caption{Detuning of the resonance frequency of the resonator as
	a function of the applied magnetic flux.
\label{fig:detuning}}
\end{figure}



We performed several measurements in a dilution refrigerator with base temperature of 10 mK, where a biasing coil was integrated into the sample holder to apply a magnetic flux.
First, we characterized the resonator by measuring its transmission and varying the input power and frequency (figure~\ref{fig:nonlinear}). The resonance curve exhibits typical behavior of a Duffing
oscillator.\cite{Duffing} From the low input power spectra, we extract the resonance
frequency of the third harmonic and its loaded quality factor, which have the values:
$f_3=7.45$~GHz and $Q=3300$, respectively.
From the theoretical model,\cite{Yurke06} we determined the ratio between the Kerr constant $K$ and the nonlinear dissipation $\gamma_nl$, which is calculated to be -3.5 for $\Phi = 0$. This kind of analysis is not possible for  $\Phi = \Phi_0/2$, because as one can see from figure~\ref{fig:detuning}, for low input power the transmission curve is bent towards the higher frequencies and then, for higher powers, the curve bends towards lower frequencies. We can model this behavior by a current (power) dependent Kerr constant
\begin{equation}
K_n = K^{(1)}_n +K^{(2)}_n\frac{I_r^2}{I_{cqr}^2},
\end{equation}
where $K^{(2)}_n$ corresponds to the fourth order term in the Taylor series given by equation (\ref{eq:L}).
From qubit spectroscopy, we obtained the parameters that allow us to determine
$\tilde{L}$ and subsequently $K^{(1)}_3 = 3\times 10^{-3}$  and  $K^{(2)}_3 = -6\times 10^{-1}$.
We investigated the detuning of the resonance frequency at different values of the applied magnetic field (figure~\ref{fig:detuning}). We
chose two working points: magnetic flux $\Phi = 0$, where the energy gap between the qubit's energy levels is large, and $\Phi = \Phi_0/2$, which corresponds to the qubits degeneracy point with an energy gap equal to the tunnelling amplitude $\Delta$ . The detuning of the resonance frequency of the resonator is
inversely  proportional to the energy gap\cite{Grajcar05a,Omelyanchouk10} and it is maximal at $\Phi = \Phi_0/2$.

\begin{figure*}[ht]
\includegraphics[width=6cm]{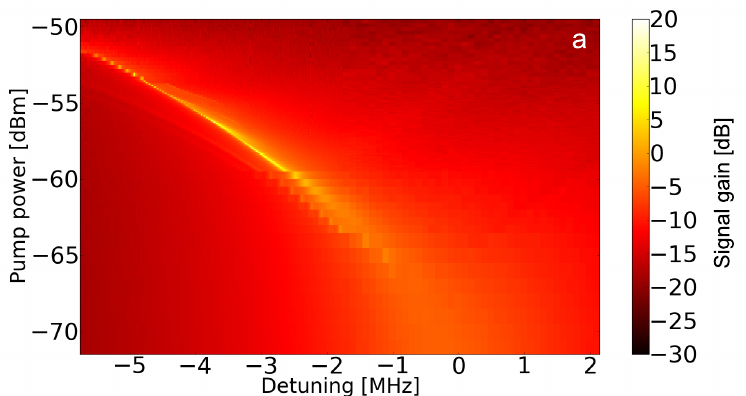}
\includegraphics[width=6cm]{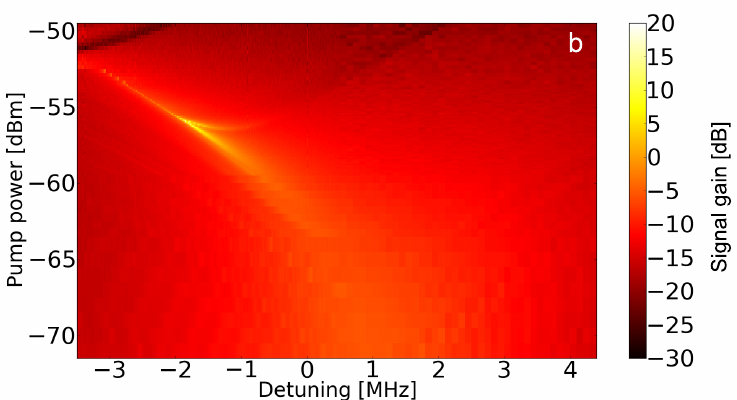}
\includegraphics[width=6cm]{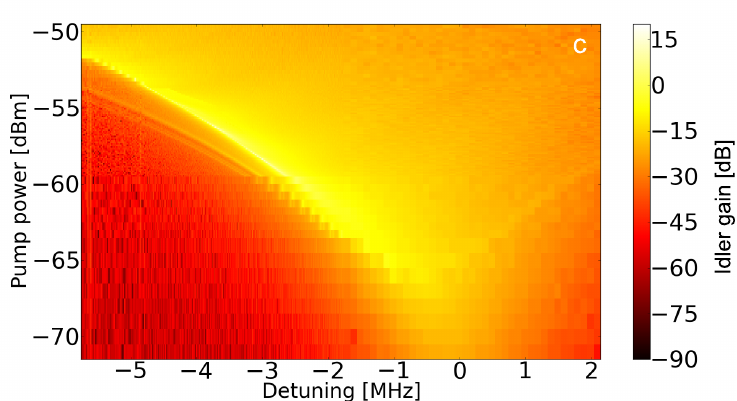}
\includegraphics[width=6cm]{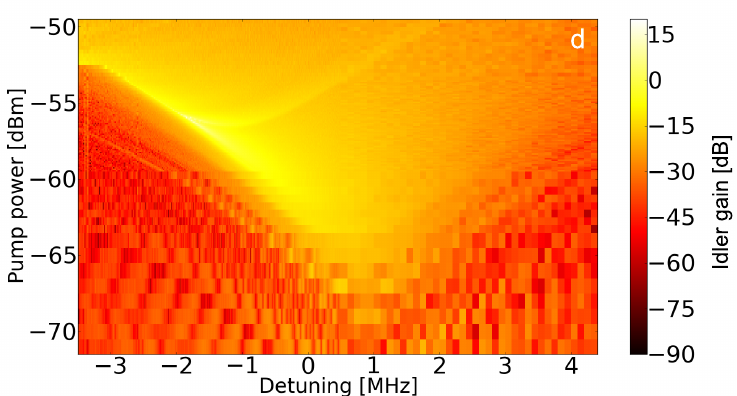}
\caption{ Gain of the parametric amplifier as a function of detuning from the resonance
frequency (x-axis) and the input pump power (y-axis). (a) Signal gain at magnetic flux $\Phi=0$  and (b) $\Phi = \Phi_0/2$ . (c) Idler gain at $\Phi=0$  and (d) $\Phi = \Phi_0/2$ .
\label{fig:gain}}
\end{figure*}

The gain of the parametric amplifier was investigated in the following way:
the transmission was measured at the signal frequency whilst keeping its amplitude constant. The
amplitude of the pump was greater than the amplitude of the signal, while simultaneously sweeping the amplitude and frequency of the pump. The
frequency difference between the signal and the pump was 10 kHz (approx. 1$\%$
of the bandwidth of our resonator).
The gain of the parametric amplifier depends on the frequency and the power of the
pump (figure~\ref{fig:gain}). The magnetic flux applied to the qubits only slightly alters the resulting gain of both the
signal and the idler, while the maximal gain is $\approx$ 20 dB. The main difference between these
two cases is the presence of two “high amplification branches” at $\Phi = \Phi_0/2$ instead of only one
present at $\Phi = 0$ and a periodic pattern of the gain depending on the bias flux (see figure~\ref{fig:gain}).
These peculiarities may be a consequence of
the quantum nature of the superconducting qubit (“Landau - Zenner beam splitting”
at the qubit degeneracy point \cite{Shevchenko08}) and/or the modulation instability
 characteristic for wave propagation in a nonlinear dispersive media.\cite{Ostrovskiy66,Agrawal87} The modulation instability has recently been investigated
in optical metamaterials taking into account both cubic and quintic nonlinearities.\cite{Saha13} It has been predicted that a combined effect of cubic–-quintic nonlinearity increases the modulation instability gain.
Our experimental results show that the quintic nonlinearity, which is characterized by
higher order Kerr constant $K^{(2)}_n$, cannot be neglected  for
strongly coupled superconducting qubits. Moreover, the shape of the idler gain is remarkably similar to the modulation instability gain shown in Refs.~\onlinecite{Agrawal87,Saha13}.
A more detailed theoretical quantum analysis of parametric amplification is required in order to clarify this effect.

\begin{table}[ht]
\caption{\label{tab:Kerr_con} Comparison of superconducting parametric amplifiers.}
\begin{ruledtabular}
\centering
\begin{tabular}{@{}cccc}
$\mathbf{K^{(1)}/\omega_0}$ & $\mathbf{K^{(2)}/\omega_0}$& \textbf{max. gain [dB]} & \textbf{reference} \\
$-1.10\times10^{-6}$ & - &30 & [\onlinecite{Eichler11}]\\
$-1.27\times10^{-9}$ & - &22 & [\onlinecite{Tholen09}]\\
$-1.6\times10^{-5}$ & - &28 & [\onlinecite{Beltran07}]\\
$+3.0\times10^{-3}$ & $-6\times10^{-1}$&20 & this paper\\
\end{tabular}
\end{ruledtabular}
\end{table}

In summary, we designed a parametric amplifier based on a superconducting coplanar
waveguide resonator with a pair of integrated qubits serving as the nonlinear element. In contrast to other publications (see table,\ref{tab:Kerr_con}) the Kerr constant of our system is $sign$-tunable and several orders of magnitude higher.
Our amplifier achieves a maximal gain of 20 dB, which is comparable to the gain of similar superconducting parametric amplifiers.\cite{Eom12,Tholen07,Abdo09} However, the experiments reveal two unknown features, namely two “branches” of high amplification of the idler signal and
in the case of a bias flux of $\Phi = \Phi_0/2$
a periodical pattern of the gain dependent on the flux itself.


The research leading to these results has received funding from the European Community’s
Seventh Framework Programme (FP7/2007-2013) under Grant No. 270843
(iQIT). This work was also supported by the Slovak Research and Development Agency
under the contract APVV-0515-10 (former projects No. VVCE-0058-07, APVV-0432-07)
and LPP-0159-09. The authors gratefully acknowledge the financial support of the EU
through the ERDF OP R$\&$D, Project CE QUTE $\&$ metaQUTE.

\def\urlprefix{}
   \def\url#1{}

\end{document}